\documentstyle[aps,epsfig,preprint]{revtex}

\begin{document}

\title{
Invaded Cluster Dynamics for Frustrated Models
}
\author{
Giancarlo Franzese, Vittorio Cataudella and Antonio Coniglio$^*$\\
Dipartimento di Scienze Fisiche, Universit\`a ``Federico II",\\ 
Mostra d'Oltremare Pad.19 I-80125 Napoli, Italy,\\
INFM - unit\`a di Napoli and \\
$^*$ INFN - sezione di Napoli\\
}

\date{\today}

\maketitle

\begin{abstract}
The Invaded Cluster (IC) 
dynamics introduced by Machta et al. [Phys. Rev. Lett.
75 2792 (1995)] is extended to the fully frustrated Ising model on a square
lattice. The properties of the dynamics which exhibits numerical evidence 
of self-organized criticality are studied. The fluctuations 
in the IC dynamics are shown to be intrinsic of the algorithm and the 
fluctuation-dissipation theorem is no more valid.
The relaxation time is found very short and does not present
critical size dependence. 
\end{abstract}

\hfill  {Ref-no. DSF-T-30/97}

\newpage

\section{Introduction}

Recently a new type of cluster Monte Carlo (MC) dynamics, the Invaded
Cluster (IC) dynamics, based on invaded  percolation
has been introduced by Machta et al. for the ferromagnetic Ising 
model.\cite{machta1}  The IC dynamics is based on the Kastelein-Fortuin 
and Coniglio-Klein (KF-CK) cluster formulation of the Ising model \cite{kf-ck}
and has been shown to be
even more efficient than the Swendsen-Wang (SW) dynamics \cite{sw}
in equilibrating the system at the critical temperature. 
The IC dynamics has the 
advantage that the value of the critical temperature does not need to be
known {\it a priori}. In fact the dynamics itself drives the system to
the critical region as in  self-organized critical (SOC) systems. 

The aim of this paper is to extend the IC dynamics to frustrated systems where 
KF-CK clusters percolate at a temperature $T_p$, higher than the critical 
temperature $T_c$. In particular we will consider the fully frustrated (FF)  
Ising model on a square lattice where it has been shown numerically 
\cite{vittorio} that $k_BT_p/J\simeq 1.69$ ($k_B$ is the Boltzmann constant and 
$J$ is the strength of the interaction) and $T_c=0$. We will use
two definitions of clusters. The first definition is based on the 
KF-CK clusters. In this case the IC dynamics leads to a SOC percolating state 
at temperature $T_p$ corresponding to the percolation of KF-CK clusters.
The second definition is more general \cite{prl} and reduces to the cluster of 
Kandel, Ben-Av and Domany (KBD) \cite{kbd} in the FF Ising model. In this case 
the IC dynamics leads to a SOC state at the thermodynamical 
critical temperature $T_c=0$.

In sec. II we review the definition and the results of IC dynamics on 
ferromagnetic Ising model. Then we extend the IC dynamics to the FF Ising model 
using the KF-CK cluster in sec. III and the KBD clusters in sec. IV where
we study also the equilibrium relaxation of the proposed dynamics.

\section{IC dynamics for ferromagnetic Ising model}

The rules which define the IC dynamics for the ferromagnetic Ising model
are  very simple. Let us start from a given
spin configuration. As first step all the pairs of nearest 
neighbour (nn) spins are ordered randomly. Then, following the
random order, a bond is activated between the nn spin pairs
only if the two spins satisfy the interaction. The set of activated
bonds partitions the lattice into clusters  which are classified every
time a new bond is activated. When one of the clusters spans
the system the procedure is stopped and the spins belonging
to each cluster are reversed all together with probability $1/2$.
The spin configuration obtained
can be used as starting point for the next application of the dynamical rule.
Successively, the algorithm has been generalized to
different stopping rules \cite{machta2,glotzer} and to different model 
(the Potts model \cite{machta2} and the Widom-Rowlinson fluid \cite{machta3}).

The rational which is behind this rule is the well known mapping between
the ferromagnetic Ising model and the correlated bond percolation.
\cite{kf-ck} In this framework bonds are introduced in the ferromagnetic
Ising model between parallel nn pairs of spins with probability
$p=1-\exp(-2\beta J)$ where $\beta=1/(k_B T)$. The clusters (KF-CK clusters),
defined as maximal set of connected bonds, represent sets of correlated spins 
and percolate exactly at the critical temperature, $T_p=T_c$. 
The well known SW cluster dynamics \cite{sw} uses these clusters to sample very
efficiently the phase space at any temperature. In this dynamics at each 
MC step the KF-CK clusters are constructed and the spins belonging
to each cluster are reversed all together with probability $1/2$.
This produces a new spin
configuration on which a new cluster configuration can be build up.
The IC dynamics is very similar to SW dynamics and differs only
on the way clusters are constructed.
Since the dynamic
rule introduced by Machta et al. builds up clusters which, by definition,
percolate through the system, it is expected that the average properties
of the clusters are the same as the KF-CK clusters at the critical
temperature. In fact, in Ref.\cite{machta1} it has been shown that
the ratio of activated bond to satisfied interactions,
in the large $L$ limit ($L$ is the linear lattice size), 
is very close to the critical probability
$p_c=1-\exp(-2J/(k_BT_c))$ with which KF-CK clusters are constructed
and it has a vanishing standard deviation. 
Furthermore, the estimated mean energy also tends 
in the large $L$ limit
to a value very close to the critical  equilibrium value $E(T_c)$
with the finite size behaviour expected in the ferromagnetic Ising model
at the critical point.
On the other hand the energy fluctuation $C=\langle E^2\rangle -
\langle E\rangle^2$ is not related to the specific heath.
In fact, it has been found
that $C$ diverges linearly with $L$ and not logarithmically as the
specific heat does. 
The latter result points to the fact that the IC dynamics does not
sample the canonical ensemble in finite volume and, therefore,
energy fluctuations and specific heat are not any longer related
by the fluctuation-dissipation theorem. Since the fluctuation
$C^{1/2}/L^2 \mapsto 0$ for $L \mapsto \infty$, it is generally assumed 
that in the thermodynamic limit the IC dynamics is equivalent
to the canonical ensemble even if a rigorous proof is still lacking.

\section{IC dynamics for FF Ising model with KF-CK clusters}

We now extend the IC dynamics to 
frustrated systems where the KF-CK clusters percolate at a temperature
$T_p$ which is higher than the critical one $T_c$. In particular we will 
consider the FF Ising model on square lattice where \cite{nota0} 
$T_p\simeq 1.69$ 
\cite{vittorio} and $T_c=0$. The FF Ising model is defined by the Hamiltonian
\begin{equation}
H =  -J \sum_{\langle ij \rangle} (\epsilon_{i,j} S_i S_j - 1)
\label{h}
\end{equation}
where $S_i$ takes the values $\pm 1$ and $\epsilon_{i,j}$ assumes the value 
$-1$ on even columns and is $+1$ otherwise.
 
In the simplest extension of the IC dynamics we introduce bonds at random between 
spins satisfying the interaction (i.e. $\epsilon_{ij}S_iS_j=1$) until a 
spanning cluster is found. As in the ferromagnetic case, at this point the 
spins belonging to each cluster are reversed all together with probability 
$1/2$. The procedure is iterated until equilibrium is reached. We have done
simulations performing 
measurements over $5 \cdot 10^3 $ MC sweeps after discarding the first 
$10^3$ for equilibration for systems with sizes $L$ ranging from 16 to 300, 
and over $3.75 \cdot 10^3 $ MC 
sweeps after discarding the first 750 for $L=350$. 

In Fig.1.a we show as functions of the system size $L$
the ratio of activated bonds $N_b$ to satisfied interaction
$N_s=2L^2-E/2$ (where $E$ is the energy), $\langle f\rangle=N_b/N_s$,
and in Fig.1.b the density of energy $\epsilon=E/L^2$. 
Assuming the size dependences 
$\langle f\rangle= \langle f\rangle_{\infty} - A/L$ and 
$\epsilon = \epsilon_{\infty} + A/\log(L)$ we find
$\langle f\rangle_{\infty}=0.698 \pm 0.008$ and 
$\epsilon_{\infty} =1.24\pm 0.01$, which are very close to the values 
$p_c=1-e^{-2/T_p}\simeq 0.694$ and 
$\epsilon(T_p)=1.234$ at the temperature $T_p\simeq 1.69$.
The errors on $\langle f\rangle$ and $\epsilon$ decrease with increasing $L$, 
denoting that the $\langle f\rangle$ and the $\epsilon$ distributions become
sharp when $L\mapsto\infty$ .

The results obtained show that the dynamics has not driven the 
system into the critical thermodynamical state at $T=0$, but to the 
percolation critical state at $T=T_p$.
In order to study the convergence as function of $L$ 
we have extracted for the IC dynamics at each $L$ an effective 
temperature $T_{IC}(L)$ (see Tab.1) by using 
$\langle f\rangle=1-\exp(-2/T_{IC})$, then we have compared the IC energy 
density to the analytical energy density 
$\epsilon(T_{IC})$ at $T_{IC}$ (Fig.1.b). 
This analysis clearly show that the $L\mapsto\infty$ limit is reached in the 
IC dynamics very slowly. In particular the energy is systematically larger 
than that obtained by SW dynamics. 
Furthermore, since the dynamics build up, by definition, percolating clusters 
the mean cluster size $S/L^2$ of IC clusters diverges in the following way
$S/L^2 = A L^{\gamma_p / \nu_p}$ where $\gamma_p / \nu_p = 1.78 \pm 0.11$ is
a good approximation of the exponent $\gamma_p / \nu_p = 1.792$ 
of Random Bond Percolation \cite{Stauffer_Aarony} and $A=0.022 \pm
0.013$ (Fig.2). This result is in excellent agreement with the behaviour 
found for KF-CK clusters for both the critical exponent
and the prefactor $A$.\cite{vittorio} This critical behaviour drives also 
energy and magnetization fluctuations in a critical regime, i.e.,
$\langle E^2\rangle - \langle E\rangle^2 \sim L^{2.9}$ and 
$\langle M^2\rangle - \langle M\rangle^2 \sim L^{3.8}$ (Fig.3). This result is 
in strong contrast with the behaviour of specific heat and magnetic
susceptibility at $T=T_p$ which behave as $L^2$. 
In fact $T_p$ is not a critical thermodynamic
temperature and both specific heat and magnetic
susceptibility are finite at $T=T_p$. This again stresses the fact that 
the IC dynamics does not sample the canonical ensemble in finite volume
and shows that the spin configurations visited by the IC
dynamics with a finite probability correspond to a
much larger range of energies and magnetization than
in any ordinary dynamics.

It is interesting to note that the straightforward
application of the IC dynamics to more complex systems as the Ising
spin glass (SG) model could allow to sample very efficiently the
equilibrium spin configurations of the system at $T=T_p$
where the KF-CK clusters percolate (in 2d SG $T_p\simeq 1.8$ \cite{vittorio}, 
in 3d SG $T_p\simeq 3.95$ \cite{lucilla})
with a dynamics which exhibits SOC. However since $T_p$ is usually much larger 
than the spin glass critical temperature $T_{SG}$ (in 2d $T_{SG}=0$, in 3d 
$T_{SG}\simeq 1.11$ \cite{KawaYoung}), any other dynamics does not 
suffer for slowing down. 

\section{IC dynamics for FF Ising model with KBD clusters} 

We ask now how to build up an IC dynamics able to drive a system to the 
frustrated thermodynamical critical temperature. 
From the analysis and the argument given for the ferromagnetic
case we understand that the cluster definition needs to be modified in such a 
way that $i)$ the frustrated system can still be mapped on the corresponding correlated 
percolation model and $ii)$ the clusters percolate at the thermodynamical 
critical temperature. \cite{nota1} The first conditions is always satisfied 
by the KF-CK clusters  \cite{coniglio_mapping_sg} but not the second one.
A general procedure to construct such clusters in a systematic way was 
suggested in Ref.\cite{prl}. In particular for the FF model this procedure 
leads to the cluster algorithm \cite{kbd} proposed by Kandel, Ben-Av and Domany 
(KBD). \cite{nota2}
To define the clusters in KBD dynamics one
partitions the square lattice in a checkerboard way and chooses randomly one 
of the two patterns (Fig.4a). For each plaquette if three 
of four spin pairs are satisfied (a single plaquette in the FF model can have 
either one or three satisfied spin pairs) one activates bonds between 
the two spin pairs satisfying the interaction and facing each other 
(Fig.4b) with a probability $p=1-\exp(-4/T)$. 
Numerically it was showed \cite{giancarlo}
that the clusters obtained with such procedure 
percolates at the thermodynamical critical temperature $T_c=0$, with critical 
exponents $\nu_p= 1$ and $\gamma_p= 2$ to be compared with the thermodynamical 
critical exponents $\nu=1$ and $\gamma=3/2$.

With this idea in mind we propose the following invaded cluster dynamics. 
From the checkerboard partition we order randomly all the square
plaquettes belonging to the chosen pattern.  
Then plaquettes are tested in this order to see how many spin pairs are 
satisfied and, in plaquettes with three spin pairs satisfying the interaction,
we activate bonds between the two spin pairs satisfying the interaction and 
facing each other.
Every time a pair of bonds is activated the cluster structure change
and the occurring of a spanning cluster is checked. As in the 
previous cases when the first cluster percolates the
cluster evolution is stopped and  a new spin configuration is
obtained by reversing the spin belonging to each cluster
all together with probability $1/2$.
The dynamics proposed is related to the
algorithm introduced by KBD exactly as the IC
dynamics by Machta et al. is related to the SW algorithm. 

We have tested the proposed algorithm performing measurements over $10^4$ MC
sweeps on square lattices of sizes $L$ ranging from $16$ to $200$ and over 
$5 \cdot 10^3$ MC sweeps on $L= 250 \div 400$, 
after discarding the first $10^3$ for equilibration.
 
In Fig.5.a  we show the results of our
simulations which give a situation very similar to that obtained 
in the ferromagnetic case. In fact, the ratio  of plaquettes with 
activated bonds $N_b$ to plaquettes with three satisfied interactions
$N_s=(3L^2-E)/4$, $\langle f \rangle=N_b/N_s$,  reaches 
the value $p_c =1 $ for large  system size $L$.
At the same time the effective temperature 
$T_{IC}$ (see Tab.2), obtained by $\langle f \rangle =1-\exp(-4/T_{IC})$, 
converges rapidly to the limit value $T_c=0$. In Fig.5.a we shows also 
$p_p=1-\exp(-4/T_p(L))$ where $T_p(L)$ is the estimated \cite{giancarlo} 
percolation temperature of KBD clusters in a square FF system of size $L$.
The mean energy per spin (Fig.5.b) $\epsilon$ tends, within the 
numerical precision obtained, towards the thermodynamic value
at the critical temperature $\epsilon(T=0) =1$ with the expected size 
dependence $\epsilon-\epsilon_{\infty} = AL^{-1} + BL^{-2}$. 
The fit of the data gives $\epsilon_{\infty} =1.016\pm 0.002$. 
%The same behaviour is found for the number
%of the bonds per spin $n_b$ (Fig.?). In this case the fit gives a value
%$n_b(T=0)=0.99(2)$ even closer to the exact value $n_b=1$. 
The errors on $\langle f \rangle$ and $\epsilon$ go to zero with increasing 
$L$. The estimated mean cluster size exponents
$\gamma_p/\nu_p=1.2 \pm 0.2$ (Fig.6) does not coincide with the expected values 
$\gamma_p/\nu_p=2$.\cite{giancarlo} 
We explain this result with the very slow convergence as function of $L$ of 
the percolation quantities obtained with the KBD-clusters
(see for example Fig.8 of Ref.\cite{giancarlo}). Thus we expect to recover 
the right behavior only for sizes fairly large.
We have studied the energy and magnetization fluctuations also in this case
(Fig.7) and found that 
$\langle E^2\rangle - \langle E\rangle^2 \sim L^{2.1}$ and 
$\langle M^2\rangle - \langle M\rangle^2 \sim L^{3.1}$.
These exponents, as in the previous cases do not
agree with those  expected for specific heat and magnetic susceptibility.
As in the ferromagnetic case we obtain that
energy and magnetization fluctuations are larger in the IC dynamics than in 
the canonical ensemble.

We have also studied the equilibrium  relaxation of the magnetization of the 
proposed IC dynamics
\begin{equation}
\phi(t) = \frac{\langle M(t')^2 M(t'+t)^2\rangle - \langle M(t')^2 \rangle^2}
{\langle M(t')^4\rangle - \langle M(t')^2 \rangle^2}
\end{equation}
where the time $t$ is measured in MC steps. As shown in Fig.8
$\phi(t)$ vanishes in few MC steps. 
The integrated autocorrelation time $\tau$ defined as
\begin{equation}
\tau = \frac{1}{2} + \sum_{t=1}^{t_m} \phi(t)
\end{equation}
is reported  in Tab.3 for different values of $L$.  The dependence  on
$t_m$ is extremely week since $\tau$ reaches a plateau very quickly.
The value obtained $\tau\sim 1.6$ is lower than that obtained in the
KBD dynamics $\tau_{KBD}\sim 2.4$ and show a weak tendency to decrease
with increasing $L$.

\section{Conclusions}

We have shown how the IC dynamics introduced by Machta et al.
for the ferromagnetic Ising model can be extended to the FF Ising model on a 
square lattice. The straightforward extension with KF-CK clusters shows that 
the IC dynamics leads to a Self Organized Critical (SOC) percolating state at 
the percolation temperature $T_p$ and 
can be used to produce equilibrium spin configurations
at temperature different from the thermodynamical critical temperature, actually at the
percolation temperature $T_p$. 
The dynamics is characterized by intrinsic diverging fluctuations and
the fluctuation-dissipation theorem is no more valid.
The extension with KBD clusters, whose percolation point coincide in the 
large $L$ limit with the critical point of the FF system, 
has properties very similar to those obtained in the ferromagnetic model:
it drives the system to the critical region without a previous knowledge
of the critical temperature and gives a reasonably good estimation of the
average energy at the critical point. The estimated integrated autocorrelation 
time is smaller than that obtained in the KBD dynamics. We have also 
stressed  that the extension has been possible since there exists a percolation
model into which the FF square Ising model can be exactly mapped. The extension 
to other frustrated systems, such as spin glass, in principle can be done using 
the systematic procedure suggested in Ref.\cite{prl}. 
The computation have been done on DECstation 3000/500 with Alpha processor.

\newpage

%\section*{Tables}

\begin{table}[ht]
\begin{center}
\caption{KF-CK clusters: numerical estimates of $T_{IC}(L)$.[10]
}
\begin{tabular}{c|c c c c c c c c c c}
$L$         & 16  & 50  & 64  & 80  & 100 & 150  & 200  & 250  & 300  &350\\
\hline
$T_{IC}(L)$ &1.98 &1.77 &1.75 &1.73 &1.72 &1.703 &1.695 &1.689 &1.686 &1.682\\
error       &0.04 &0.02 &0.01 &0.01 &0.01 &0.007 &0.007 &0.005 &0.004 &0.004\\
\end{tabular}
\end{center}
\end{table}
 
\begin{table}[ht]
\begin{center}
\caption{KBD clusters: numerical estimates of $T_{IC}(L)$.[10]
}
\begin{tabular}{c|c c c c c c c c c c}
$L$         & 16   & 50   & 64   & 80   & 100  & 150  & 200  & 250  & 300  & 400\\
\hline
$T_{IC}(L)$ & 1.05 & 0.76 & 0.71 & 0.67 & 0.64 & 0.58 & 0.54 & 0.51 & 0.49 & 0.46\\
error       & 0.15 & 0.17 & 0.12 & 0.12 & 0.14 & 0.16 & 0.19 & 0.19 & 0.25 & 0.46\\
\end{tabular}
\end{center}
\end{table}
 
\begin{table}[ht]
\begin{center}
\caption{Integrated autocorrelation time $\tau$ for IC dynamics with KBD
clusters and $\tau_{KBD}$ for KBD dynamics
}
\begin{tabular}{c|c c c c}
$L$          & 100             & 200           & 300           & 400\\
\hline
$\tau$       &$1.74\pm 0.02$   &$1.73\pm 0.02$ &$1.67\pm 0.02$ &$1.60\pm 0.02$\\
$\tau_{KBD}$ &$2.800\pm 0.008$ &$2.48\pm 0.02$ &$2.56\pm 0.01$ &$2.37\pm 0.03$\\
\end{tabular}
\end{center}
\end{table}

\newpage

%\section*{Figure Captions}

\begin{figure}[ht]
\caption{ KF-CK clusters: ratio of activated bonds to 
satisfied interaction  $\langle f\rangle$ (a) and
energy density $\epsilon$ (b) versus system size $L$ 
for a square FF system; squares are the IC dynamics output
and the superimpose lines are the plot of the fit functions
$\langle f\rangle=\langle f\rangle_{\infty}-A/L$
(with $\langle f\rangle_{\infty}=0.698\pm 0.008$ and $A=1.2 \pm 0.2$) and 
$\epsilon = \epsilon_{\infty} + A/\log(L)$ (with $\epsilon_{\infty} =1.24\pm
0.01$ and $A=0.62\pm 0.07$) for $L\geq 100$;
circles are the energy density values of a standard MC 
dynamics at the estimated temperature $T_{IC}$ in Tab.1; 
dashed line is the analytical energy density for the model at 
$T_{IC}$; 
dotted line is the asymptotic value at $T=1.69\simeq T_p$;
arrows show the asymptotic values estimated by the fits;
where not shown the errors are included in the symbols; 
all quantities are dimensionless.[10]
}
\end{figure}

\begin{figure}[ht]
\caption{ KF-CK clusters: mean cluster size per spin $S/L^2$ for IC dynamics 
for a square FF system with $L= 16 \div 350$.
We show the fit over data points for $L \geq 150$ with the best fit parameters 
given in the text ($A=0.022 \pm 0.013$ and $\gamma_p / \nu_p = 1.78
\pm 0.11$);
the errors are included in the symbols;
all quantities are dimensionless.[10]
}
\end{figure}

\begin{figure}[ht]
\caption{ KF-CK clusters: density of energy fluctuation (a) and of 
magnetization fluctuation (b) 
for a square FF system; squares are the fluctuations for IC dynamics;  
circles are the fluctuation values of a standard MC
dynamics at the estimated temperature 
$T_{IC}$ in Tab.1; 
dotted line is the analytical values for the model at $T_{IC}$; 
the estimated behaviour are 
$\langle E^2\rangle - \langle E\rangle^2/L^2 \sim L^{0.9}$ and 
$\langle M^2\rangle - \langle M\rangle^2/L^2 \sim L^{1.8}$ (the error on the 
exponents is on the last given digit);
where not shown the errors are included in the symbols;
all quantities are dimensionless.[10]
}
\end{figure}

\begin{figure}[ht]
\caption{Square FF lattice: the solid (dashed) lines are ferromagnetic 
(antiferromagnetic) interactions.
(a) Checkerboard  partition of a square lattice: 
the plaquettes with (without) a dot give a pattern;
(b) an example of plaquette with three satisfied interactions: in this case 
we activate the two vertical bonds.
}
\end{figure}

\begin{figure}[ht]
\caption{KBD clusters: ratio of plaquettes with 
activated bonds to plaquettes with 3 satisfied interactions 
$\langle f \rangle$ (a) and energy density $\epsilon$ (b) versus system 
size $L$ for a square FF system; squares are the IC 
dynamics output and the superimpose lines are the plot of 
the fit functions
$\langle f\rangle=\langle f\rangle_{\infty}-A/L-B/L^2$
(with $\langle f\rangle_{\infty}=1.000 \pm 0.003$, $A=0.1\pm 1.2$ 
and $B=9\pm 100$) and 
$\epsilon=\epsilon_{\infty}-A/L-B/L^2$
(with $\epsilon_{\infty}=1.016 \pm 0.002$, $A=4.3\pm 0.5$ 
and $B=1.0\pm 0.7$) for $L\geq 100$; 
stars are the values of $p_p=1-\exp(-4/T_p(L))$ with $T_p(L)$ 
percolation temperature of KBD clusters in a square FF system of size $L$
(from Ref.[17]); 
circles are the energy density values of KBD dynamics 
at the estimated temperature $T_{IC}$ in Tab.2 (the errors are asymmetric 
because they are derived from the indetermination on $T_{IC}$);
dashed line is the analytical energy density for the model at
$T_{IC}$;
dotted line is the asymptotic value at $T_c=0$;
arrow shows the asymptotic value estimated by the fit;
where not shown the errors are included in the symbols; 
all quantities are dimensionless.[10]
}
\end{figure}

\begin{figure}[ht]
\caption{ KBD clusters: mean cluster size per spin $S/L^2$ for 
IC dynamics for a square FF system with $L= 16 \div 400$.
We show the linear fit $\ln(S/L^2)=A+\gamma_p / \nu_p \ln(L)$ over data 
points for $L \geq 250$ with $A=-0.8\pm 1.5$ and $\gamma_p / \nu_p = 1.2 
\pm 0.2$;
the errors are included in the symbols; 
all quantities are dimensionless.[10]
}
\end{figure}

\begin{figure}[ht]
\caption{ KBD clusters: the density of energy fluctuation (a) and of 
magnetization fluctuation (b) 
for a square FF system; squares are the fluctuations for IC dynamics;  
circles are the fluctuation values of a standard MC
dynamics at the estimated temperature 
$T_{IC}$ in Tab.2 (the error are asymmetric 
because are derived from the indetermination on $T_{IC}$); 
dashed line is the analytical values for the model at $T_{IC}$; 
the estimated behaviour are 
$\langle E^2\rangle - \langle E\rangle^2/L^2 \sim L^{0.1}$ and 
$\langle M^2\rangle - \langle M\rangle^2/L^2 \sim L^{1.1}$ 
(the error on the 
exponents is on the last given digit); 
where not shown the errors are included in the symbols; 
all quantities are dimensionless.[10]
}
\end{figure}

\begin{figure}[ht]
\caption{Magnetization correlation functions in a square FF system for the 
KBD dynamics (for $L=100$ at $T\simeq T_p(L)\simeq 0.45$ and for $L=400$ at 
$T=T_p\simeq 0.342$ [17]) and for IC dynamics with  KBD 
clusters (for $L=100$ and $L=400$). Time $t$ is measured in MC steps.
}
\end{figure}

\begin{figure}[ht]
\begin{center}
\mbox{\epsfig{file=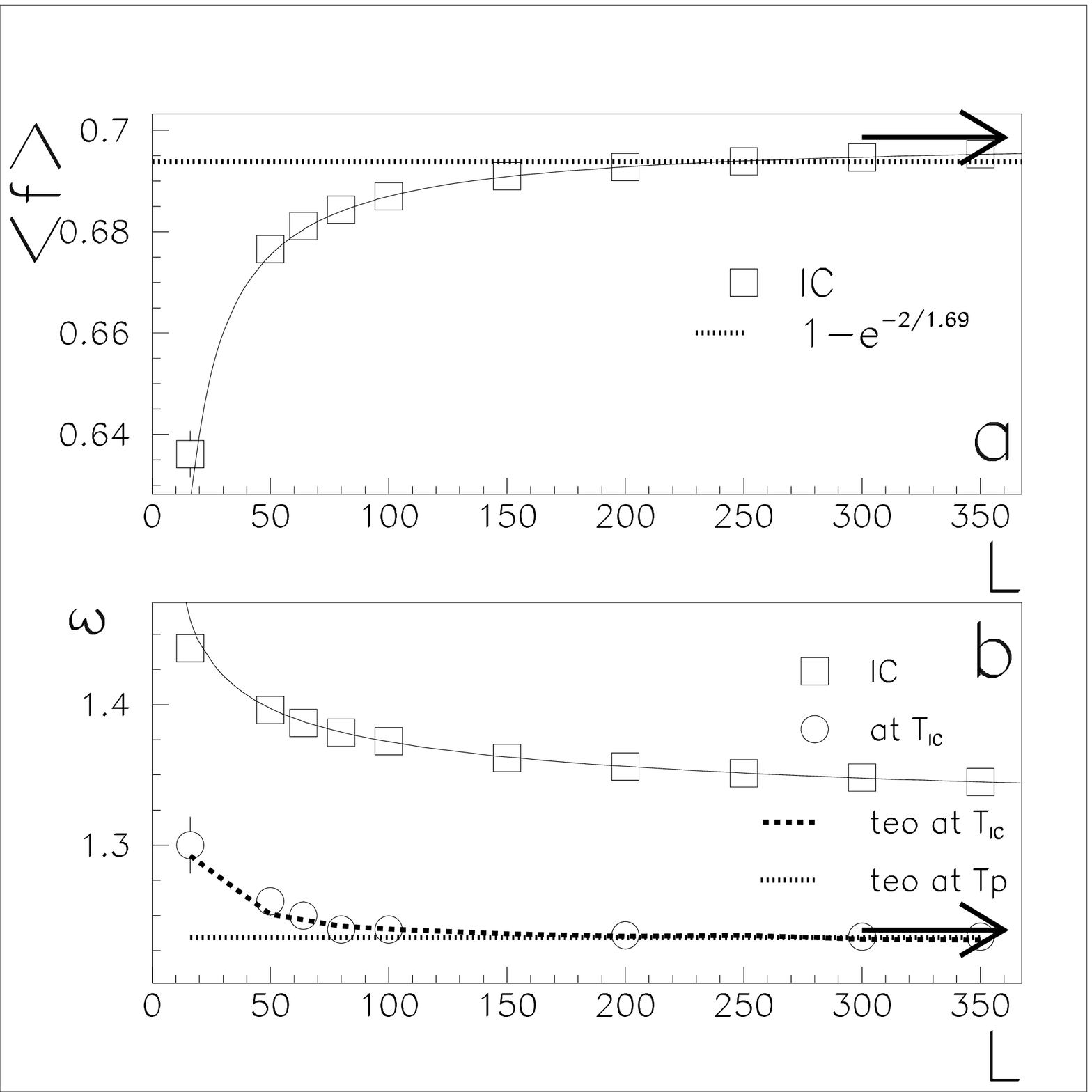,height=15.cm}}
\end{center}
\end{figure}

\begin{center}
\vfill{ FIG.1}
\end{center}

\begin{figure}[ht]
\begin{center}
\mbox{\epsfig{file=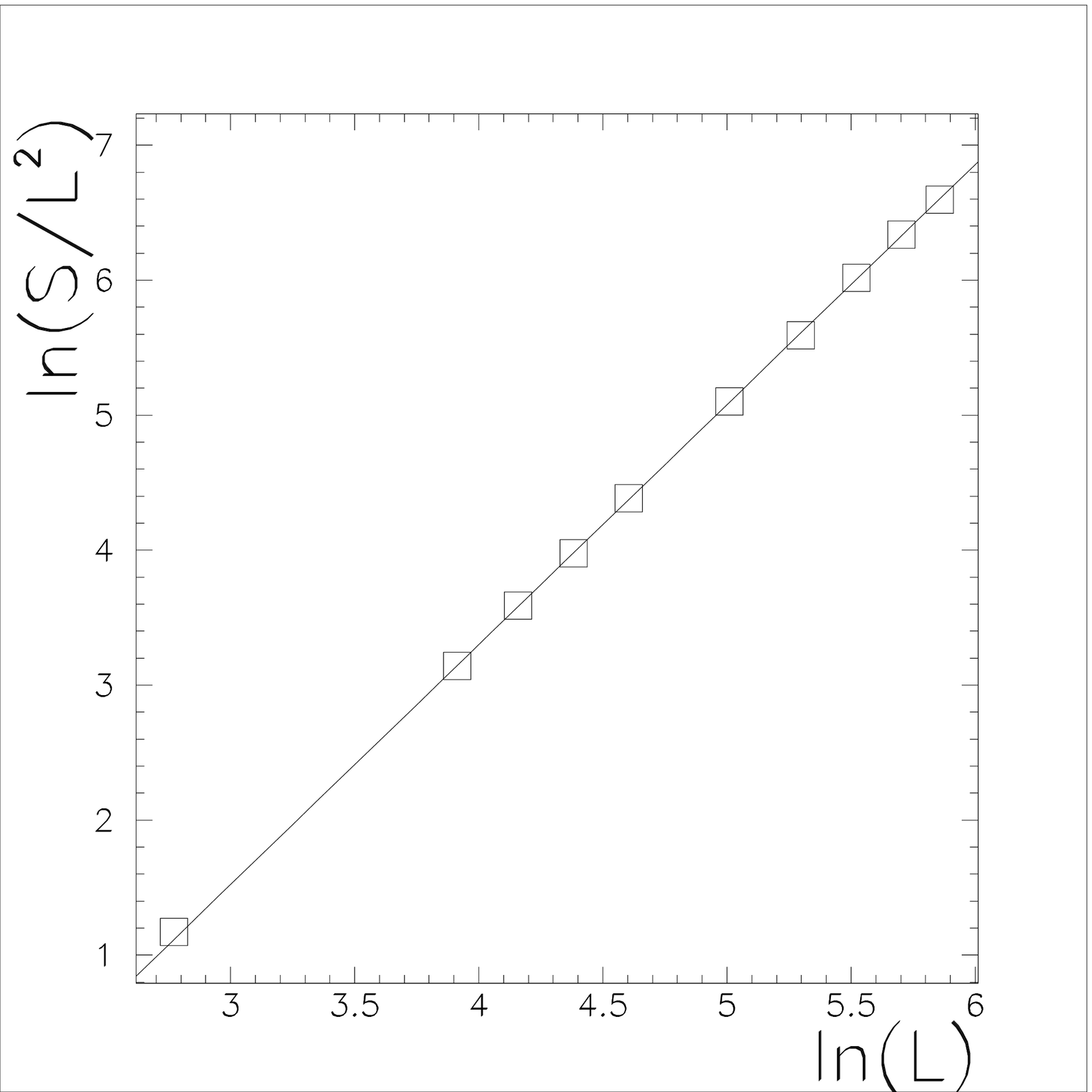,height=15.cm}}
\end{center}
\end{figure}

\begin{center}
\vfill{ FIG.2}
\end{center}

\begin{figure}[ht]
\begin{center}
\mbox{\epsfig{file=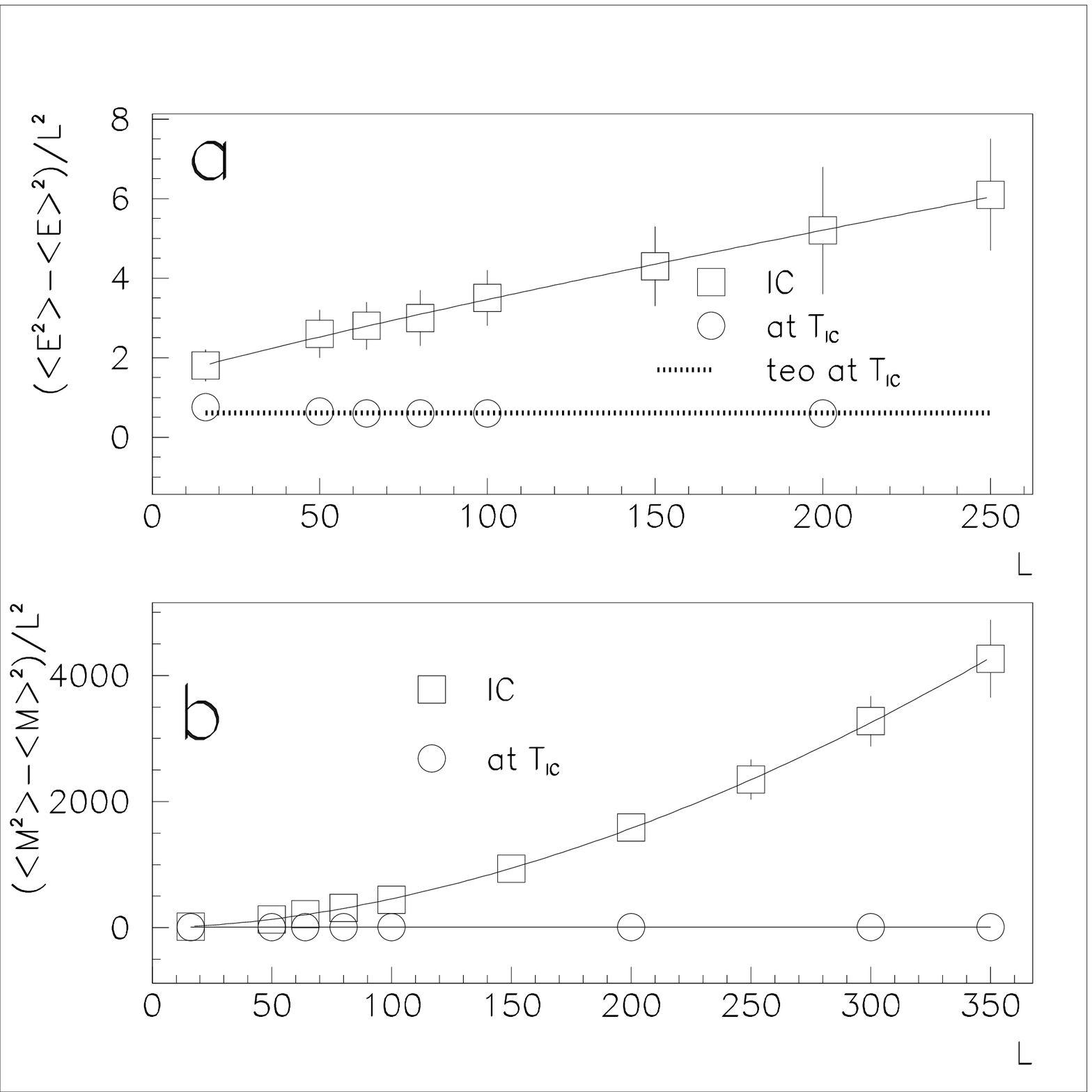,height=15.cm}}
\end{center}
\end{figure}

\begin{center}
\vfill{ FIG.3}
\end{center}

\newpage

\begin{figure}[ht]
\begin{center}
\mbox{\epsfig{file=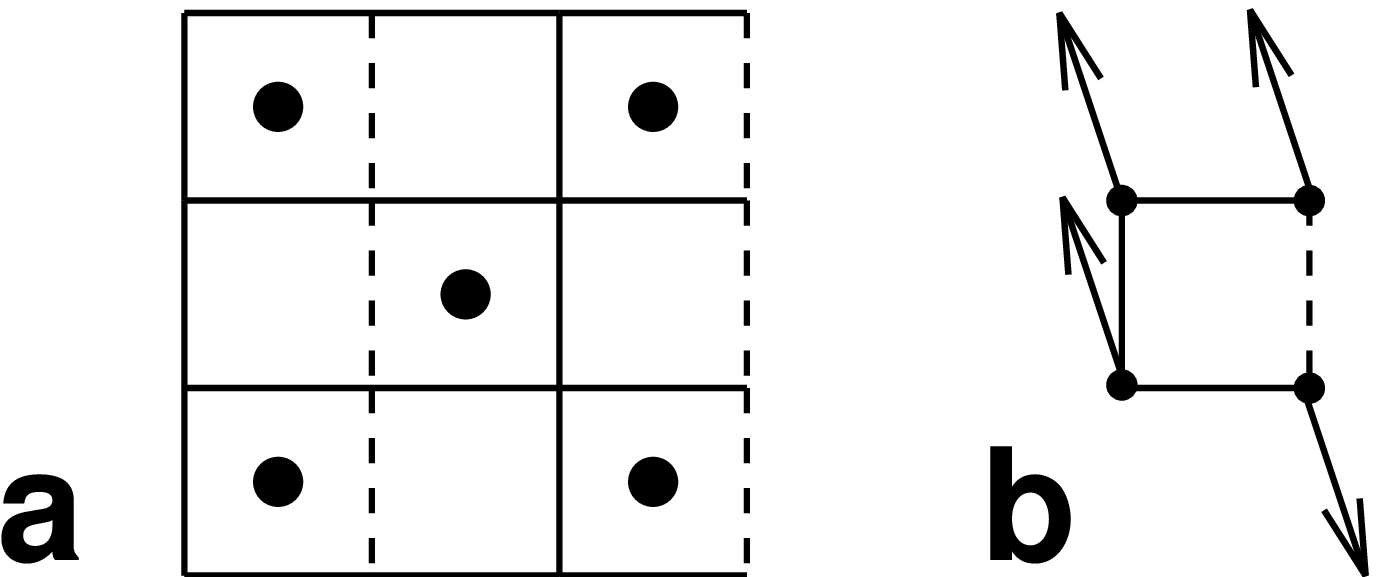,height=7.cm,width=15.cm}}
\end{center}
\end{figure}

\begin{center}
\vfill{ FIG.4}
\end{center}

\newpage

\begin{figure}[ht]
\begin{center}
\mbox{\epsfig{file=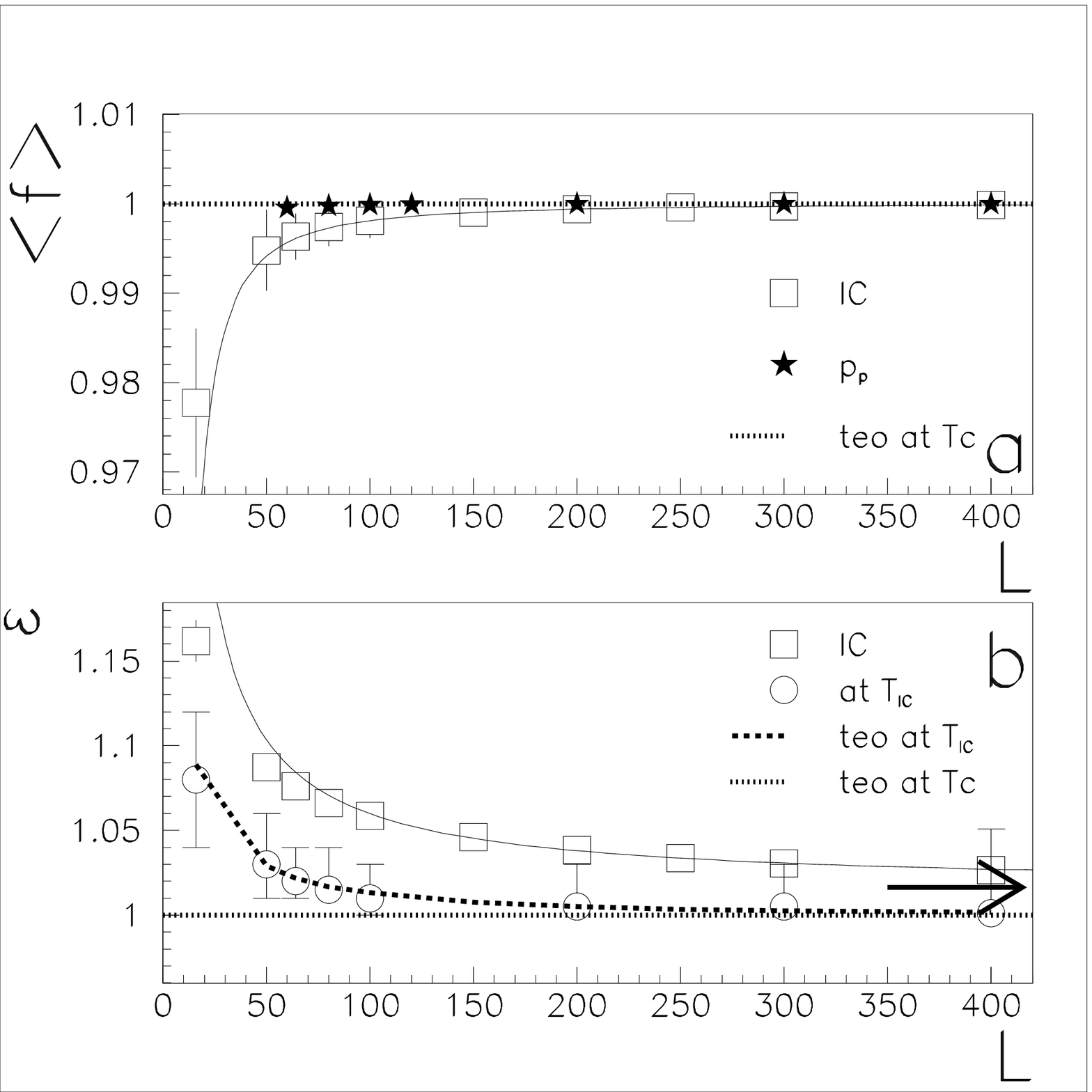,height=15.cm}}
\end{center}
\end{figure}

\begin{center}
\vfill{ FIG.5}
\end{center}

\begin{figure}[ht]
\begin{center}
\mbox{\epsfig{file=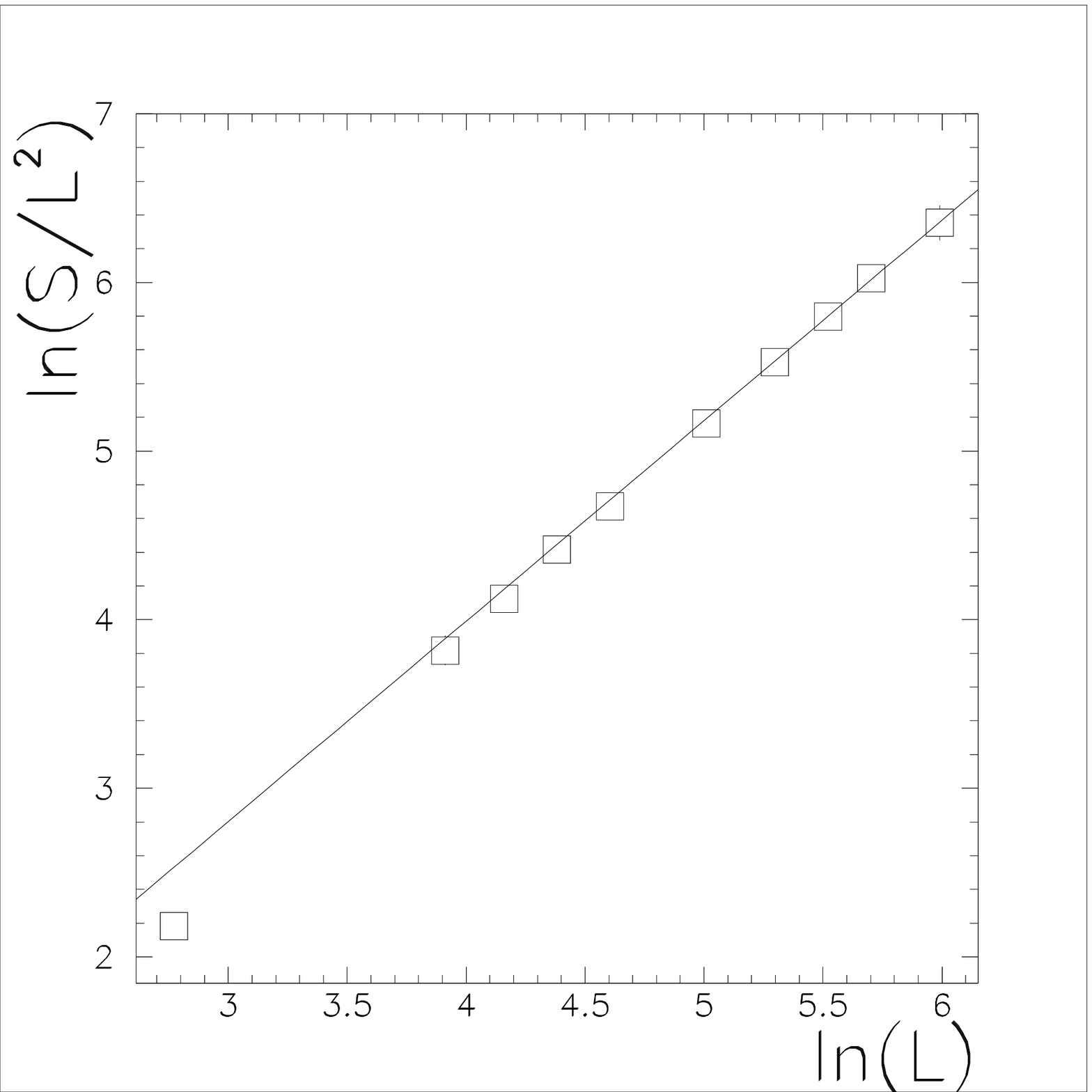,height=15.cm}}
\end{center}
\end{figure}

\begin{center}
\vfill{ FIG.6}
\end{center}

\begin{figure}[ht]
\begin{center}
\mbox{\epsfig{file=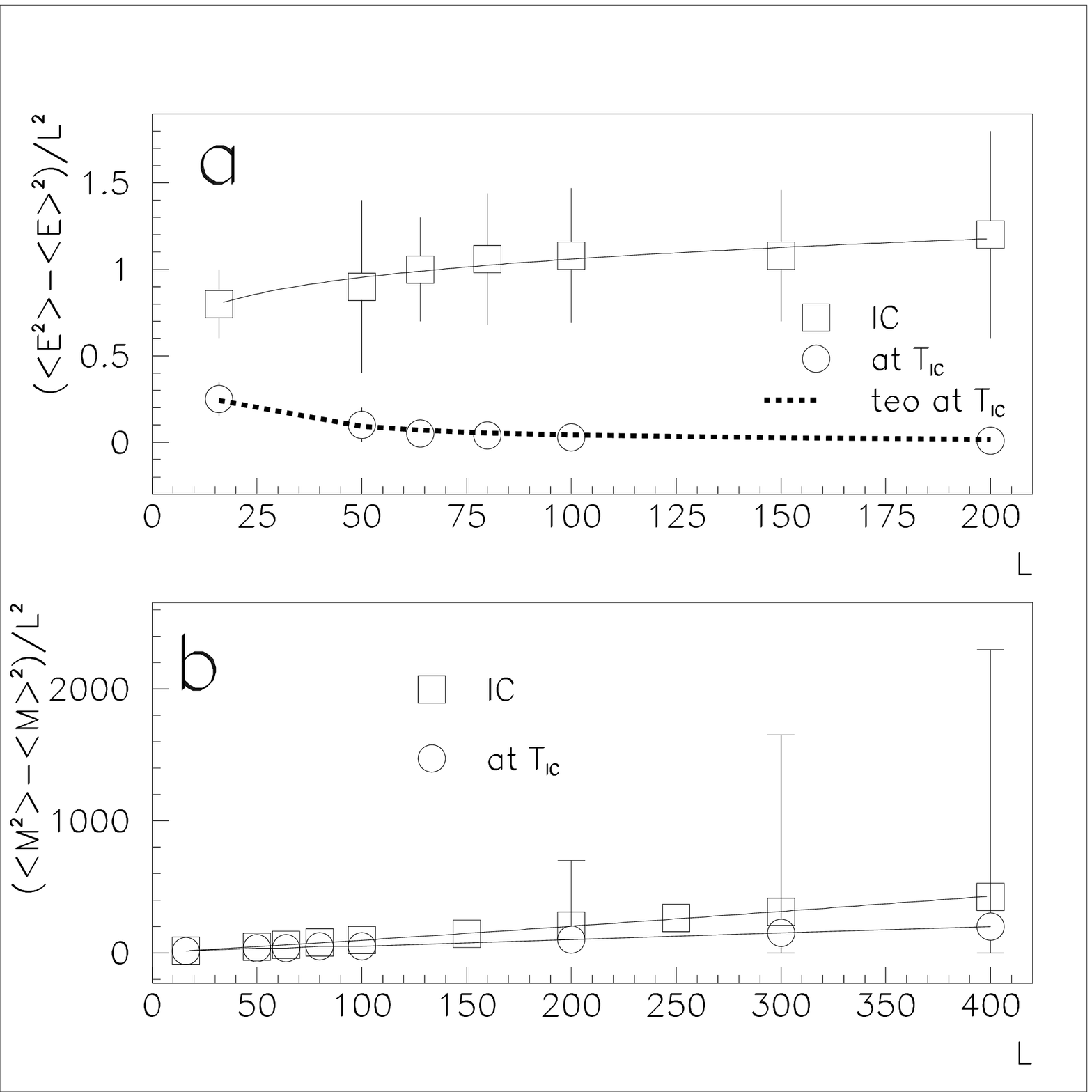,height=15.cm}}
\end{center}
\end{figure}

\begin{center}
\vfill{ FIG.7}
\end{center}

\begin{figure}[ht]
\begin{center}
\mbox{\epsfig{file=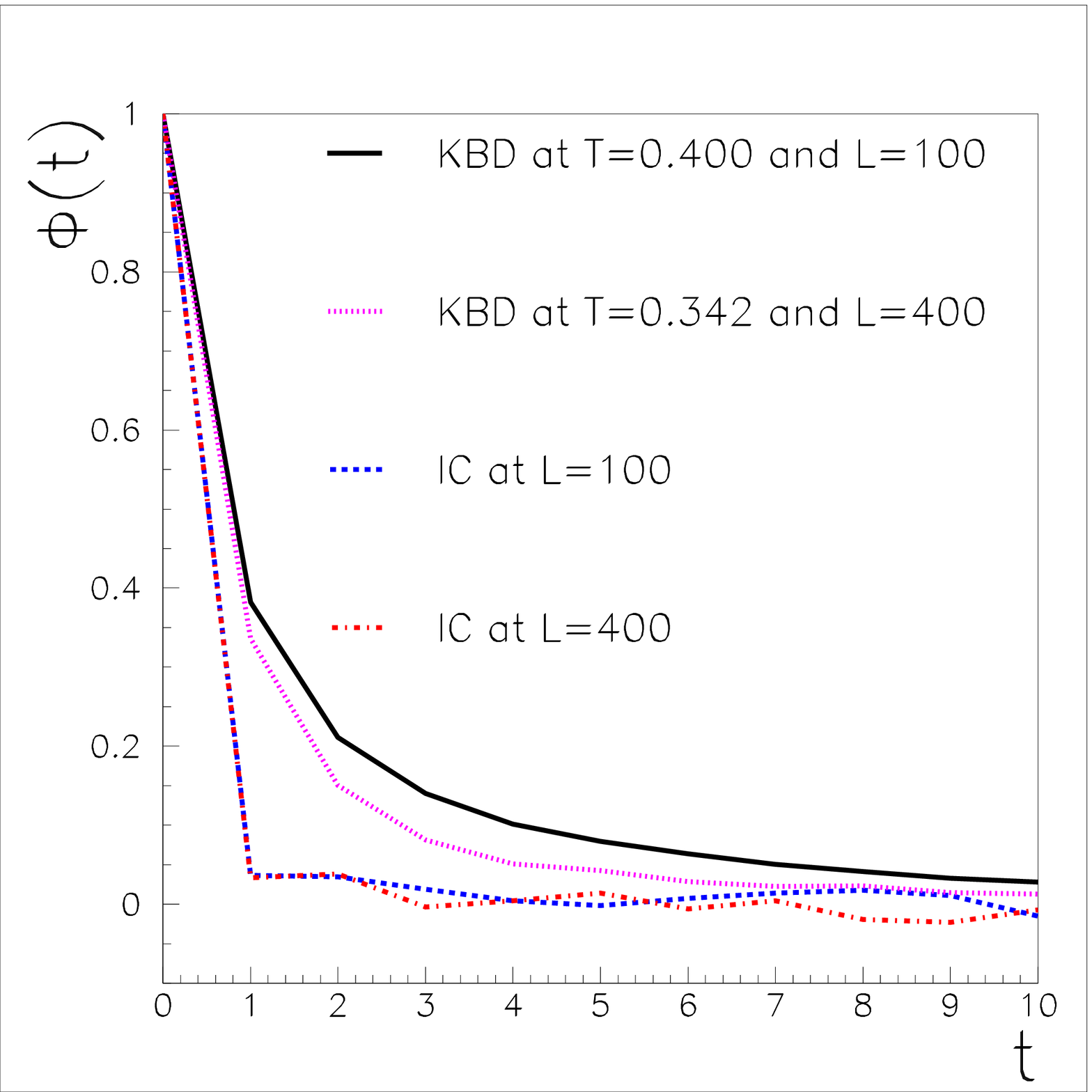,height=15.cm}}
\end{center}
\end{figure}

\begin{center}
\vfill{ FIG.8}
\end{center}


\begin{thebibliography}{99}
\bibitem{nota0} Henceforth we choose $k_B=J=1$ such that temperature 
$T=k_BT/J$ and energy $E=E/J$ are dimensionless. Furthermore we choose unitary
lattice space $a$ and unitary Bohr magneton $\mu_B$ such that the linear 
lattice size $L=L/a$ and the magnetization $M=M/\mu_B$ are also dimensionless.
In this way all considered quantities are dimensionless where not differently 
specified.
\bibitem{nota1} The two conditions $i)$ and $ii)$ have been discussed in detail 
in Ref.\cite{prl}. In general these two conditions are not sufficient to be 
sure that percolation clusters represent spin-correlated regions, i.e. to be 
sure that the set of percolation critical exponents coincides with the set of 
thermodynamical critical exponents ($\nu_p=\nu$, $\gamma_p=\gamma$ and so on).
However in many cases it has been shown \cite{prl} that these two conditions 
allow a dramatic decrease of the critical slowing down.
\bibitem{nota2} In Ref.\cite{prl} it has been also numerically 
studied a FF model (called asymmetric FF model) where the antiferromagnetic 
interaction strength is $X$ times the ferromagnetic interaction strength, with 
$0\leq X < 1$, showing that it is always possible to find a cluster definition 
(different from KF-CK one and KBD one) such that $T_p=T_c$, $\nu_p=\nu$, 
$\gamma_p=\gamma$ and so on, within the numerical precision. 
Let's note that in this paper we are considering the symmetric FF model 
($X=1$) where only $T_p=T_c$ and $\nu_p=\nu$ still holds within the numerical 
precision (see text).
\end{thebibliography}
\end{document}